\documentclass[aps,reprint,superscriptaddress]{revtex4-2}
\usepackage{graphicx}
\usepackage{xcolor}
\usepackage{amsmath,esint}
\usepackage{amssymb,bm}
\usepackage{braket}
\usepackage{enumitem}
\usepackage{float}
\usepackage{tikz}
\usepackage[colorlinks=true,linkcolor=blue,citecolor=blue]{hyperref}
\usepackage{mathtools}
\usepackage{tcolorbox}
\usepackage{relsize}
\usepackage{multirow}
\newcommand{\mbf}[1]{\textbf{#1}}

\begin{document}

\title{Testing Spontaneous Collapse Models with Coulomb Mediated Squeezing}

\author{Suroj Dey}
\email{Suroj.Dey@warwick.ac.uk}
\affiliation{Department of Physics, University of Warwick, Coventry CV4 7AL, UK.}

\author{Peter Barker}

\affiliation{Department of Physics and Astronomy, University College London,
Gower Street, London WC1E 6BT, United Kingdom}

\author{Animesh Datta}
\email{animesh.datta@warwick.ac.uk}
\affiliation{Department of Physics, University of Warwick, Coventry CV4 7AL, UK.}

\date{\today}

\begin{abstract}

We show that detecting steady-state Coulomb-mediated reduction in the thermal variance of the differential motional mode of two nanospheres can bound the Continuous Spontaneous Localization (CSL) parameter ($\lambda_{{\text{CSL}}}$). 
For realistic experimental parameters, the resulting bounds are comparable to those obtained from X-ray emission experiments and surpass those set by bulk-heating ones. Unlike these latter experiments, our bounds are robust against plausible coloured-noise extensions of collapse models. In the short-time regime, we find that a weak Coulomb-induced entanglement-based test between two charged nanospheres initialized in ground state can provide constraints on $\lambda_{\text{CSL}}$ comparable to limits set by early X-ray experiments.
\end{abstract}


\maketitle

\textit{Introduction.-} Spontaneous collapse models have been proposed to solve the measurement problem in quantum mechanics through a stochastic, non-linear modification of the Schrodinger equation, leading to the spontaneous collapse of the wave function ~\cite{RevModPhys.85.471,Bassi_2003}. These theories have an in-built amplification mechanism, such that the rate of wave-function collapse scales with the mass of the system. The mass-dependent amplification mechanism also naturally suggests that gravity plays the role of the universally coupling collapse field ~\cite{DIOSI1987377,Penrose1996,Karolyhazy1966}. Among the most prominent collapse models are the continuous spontaneous localization (CSL) model~\cite{CSL} and the Diosi-Penrose (DP) model~\cite{DIOSI1987377,Penrose1996}. The CSL model is characterized by two phenomenological parameters: The collapse rate of a single nucleon $\lambda_{\text{CSL}}$  
and the spatial resolution of the collapse process $r_{\text{CSL}}.$ 

Experimental probes of collapse models aim to constrain these parameters. The most direct approach is to observe quantum superpositions of increasingly massive systems~\cite{Arndt_2014,RevModPhys.84.157,Kaltenbaek_2023}.
However, such experiments remain technologically challenging. 
Non-interferometric tests such as spontaneous X-ray emission tests \cite{Donadi2021} and bulk heating tests \cite{Adler,Alduino2017} that target collapse-induced diffusion have  in the interim also proven highly effective in constraining collapse models, for a review of non-interferometric tests~\cite{Carlesso2022}. 

In this work, we theoretically show that detecting Coulomb-mediated squeezing in a thermal state of the differential mode in the steady state of two charged nanospheres can appreciably constrain the CSL model. By squeezing, we mean a reduction of the position variance below the thermal level.
Our set-up can improve upon bulk-heating constraints by 
several orders of magnitude for realistic experimental parameters, and even surpass the most recent X-ray emission experiment from the XENONnT collaboration~\cite{2jm3-4976} for more ambitious experimental parameters. Our constraints are also robust against plausible coloured-noise extensions of collapse models. Moreover, witnessing a weak Coulomb-mediated entanglement at short times between charged nanospheres initialized in the ground state can also constrain $\lambda_{\text{CSL}}$ appreciably. Finally, we argue that our setup cannot provide improved constraints for the Diósi–Penrose model in the near future.

Our results are based on the key insight of witnessing the squeezing of a thermal state mediated by a \textit{weak} Coulomb interaction in the face of collapse-induced diffusion. Our bounds are conservative, and a comprehensive analysis of competing noise effects will only be necessary if no squeezing is detected. More importantly, our use of two nanospheres enable independent estimation of thermal dissipation and collapse-based diffusion from their motion in the common and differential modes. At short times, the differential mode of two coupled oscillators initially cooled to the ground state is always quantum squeezed. This is of independent interest, with a potential application in testing semi-classical gravity models. Finally, we note that in this two-nanosphere set-up, collapse models predict a correlated noise, which may serve as a definitive test of collapse models.

\begin{figure}
    \centering
    \includegraphics[width=0.95\linewidth]{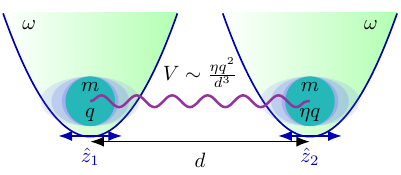}
    \caption{Two identical, levitated charged nanospheres of mass $m$ and {charges $q$ and $\eta q$, separated by a mean distance $d$. Both are harmonically trapped at angular frequency $\omega$ and coupled via $V,$ the static Coulomb interaction. 
    }}
    \label{fig1}
\end{figure}

Correlated evolution of two point-particles was known in regime of large collapse widths~\cite{Bedingham2015}. We
generalize this to macroscopic systems, allowing for arbitrary correlation lengths. Interestingly, correlated noise terms have also emerged in semi-classical Newtonian gravity~\cite{Distinguishable}, where detecting such non-local noise has been proposed as a test of classical Newtonian gravity. Since semi-classical Newtonian gravity can be formulated within the framework of spontaneous collapse of quantum matter with a feedback-induced Newtonian potential~\cite{Tilloy_2016,Tilloy_2024}, the appearance of correlated noise is expected; its experimental observation would thus constitute a test of collapse models in general.

\textit{Model and Dynamics with \text{CSL}}:
We consider two harmonically trapped masses with charges $q,$ and $\eta q$ ($\eta =\pm 1$ denotes the relative sign of the charge), masses $m$, and frequency $\omega$, separated by a {mean} distance $d$, interacting through the Coulomb potential, as in Fig.~\eqref{fig1}.
Assuming that the center of mass (COM) spreads at a scale much smaller than $d$, we expand the Coulomb potential term to up to the second order to obtain
\begin{align}
     \hat{H} = \frac{\hat{p}^{2}_{1}+\hat{p}^{2}_{2}}{2m} 
     + \frac{m\omega^{2}(\hat{z}^{2}_{1}+\hat{z}^{2}_{2})}{2} 
     + {\frac{\eta q^{2}}{4\pi\epsilon_{0}d^{3}}\left(\hat{z}_{1}-\hat{z}_{2}\right)^{2},}
\end{align}
where the canonical operators satisfy $[\hat{z}_{i},\hat{p}_{j}]=i\hbar\delta_{ij}$ for $i,j=1,2.$ In the above, we have discarded linear terms, which can always be achieved by shifting the canonical operators appropriately. They do not contribute further.
We are interested in exploring the interplay of continuous spontaneous collapse and Coulomb-mediated dynamics of two interacting massive particles. 

Restricting the COM oscillations at a scale much smaller than the collapse width ($r_{\text{CSL}}$), the evolution of the quantum state $\hat{\rho}$ under continous spontaneous localisation (CSL) can be described by (see Eq.~\eqref{lin_csl} in Appendix~\ref{app:A})
\begin{equation}
\label{eq:ME}
    \frac{\partial\hat{\rho}(t)}{\partial t} = -\frac{i}{\hbar}[\hat{H},\hat{\rho}(t)] -\sum_{\alpha,\beta\in\{1,2\}}\frac{\mathcal{D}^{\alpha\beta}_{\text{CSL}}}{\hbar^{2}}[\hat{z}_{\alpha},[\hat{z}_{\beta},\hat{\rho}(t)]],
\end{equation}
where $\mathcal{D}^{\alpha\beta}_{\text{CSL}}$ depends on the parameters $\lambda_{{\text{CSL}}}$ and $r_{\text{CSL}}.$
We also identify an \textit{correlating decoherence} term, denoting it $\mathcal{D}^{12}_{\text{CSL}}$, which decays exponentially with mean separation, see Eq.~\eqref{CSL_correlating_diffusion}.  The collapse dynamics correlate the two particles even in the absence of any Coulomb interaction, which renders the master equation non-separable.

We move to the differential (d) and common (c) modes
\begin{equation}
\hat{\mathcal{Z}}_{\text{c,d}} = 
\sqrt{\frac{m\omega}{2\hbar}}(\hat{z}_{1} \pm \hat{z}_{2}),
\hat{\mathcal{P}}_{\text{c,d}} = 
\sqrt{\frac{1}{2\hbar m\omega}}(\hat{p}_{1} \pm \hat{p}_{2}).
\end{equation}
The evolution equation for the quantum state of the differential mode, $\hat{\rho}_{\text d}(t)$, is
\begin{equation}\label{meq_d}
     \frac{\partial\hat{\rho}_{\text{d}}(t)}{\partial t} = -\frac{i}{\hbar}[\hat{H}_{\text{d}},\hat{\rho}_{\text{d}}(t)] -\frac{\mathcal{D}^{\text{d}}_{CSL}}{\hbar^{2}}[\hat{\mathcal{Z}}_{\text{d}},[\hat{\mathcal{Z}}_{\text{d}},\hat{\rho}_{\text{d}}(t)]],
\end{equation}
with
\begin{equation}
  \hat{H}_{\text{d}} = \frac{\hbar\omega\hat{\mathcal{P}}^{2}_{\text{d}}}{2 } +\frac{\hbar\omega^{2}_{\text{d}}\hat{\mathcal{Z}}^{2}_{\text{d}}}{2\omega},
  \frac{\omega_{\text{d}}}{\omega} = \sqrt{1 + \eta \delta^{2}},
  \delta^{2} = \frac{q^{2}}{\pi\epsilon_{0}m\omega^{2} d^{3}},
\end{equation}
and $\mathcal{D}^{\text{d}}_{\text{CSL}}=\mathcal{D}^{11}_{\text{CSL}}-\mathcal{D}^{12}_{\text{CSL}}.$  The master equation for the quantum state of the common mode, $\hat{\rho}_{\text{c}}(t)$, takes a similar form, with a Hamiltonian, $\hat{H}_{\text{c}}$ defined similarly to $\hat{H}_{\text{d}}$ with $\omega_{\text{c}} = \omega,$ and $\mathcal{D}^{\text{c}}_{\text{CSL}}=\mathcal{D}^{11}_{\text{CSL}}+\mathcal{D}^{12}_{\text{CSL}}.$ 

The evolution in Eq.~\eqref{meq_d} can also be equivalently described in the Heisenberg picture by the effect of a stochastic force $\zeta(t)$ with correlation: $\langle\zeta(t)\zeta(t')\rangle = \mathcal{D}_{\text{CSL}}\delta(t-t')$~\cite{Nimrichter2}.
We also incorporate generic dissipative thermal noise $\xi(t)$, which is expected to be present in any experiment. The thermal noise correlator, under Markovianity and resonance, satisfies $\langle \xi(t)\xi(t')\rangle$ = $\mathcal{D}_{\text{th}}\delta(t-t')$ where ${\mathcal{D}_{\text{th}}}= m\gamma\hbar\omega_{\text{d}} \mathcal{N}_{\text{d}},$
{$\mathcal{N}_{\text{d}} = \coth\left({\hbar\omega_{\text{d}}/{2k_{\text{b}}\text{T}}}\right),$ and} $\text{T}$ is the environmental temperature.

\textit{Steady-State Covariance Matrix.}- We assume the initial motional quantum state of the system to be in a thermal state with covariance matrix $\sigma^{(0)} = {(\mathcal{N}/2)} \mathbb{I}_{2\times2}$, where $\mathcal{N}= \coth\left({\hbar\omega/{2k_{\text{b}}T}}\right).$ The differential mode subsequently evolves with a shifted frequency $\omega_{\text{d}}$, subject to both dissipative thermal noise and diffusive collapse noise. The elements of the covariance matrix {$\sigma_{\text{d}}(t)$} in the steady-state are

\begin{align}\label{cov_matrix}
\sigma^{\infty}_{\mathcal{Z}_{\text{d}}\mathcal{Z}_{\text{d}}}&=\dfrac{\omega}{2\omega_{\text{d}}} \mathcal{N}_{\text{d}} 
+ \dfrac{\mathcal{D}^{\text{d}}_{\text{CSL}} \, \omega}{2 m \gamma \hbar \omega_{\text{d}}^{2}},\\
\sigma^{\infty}_{\mathcal{P}_{\text{d}}\mathcal{P}_{\text{d}}}&=\dfrac{\omega_{\text{d}}}{2\omega} \mathcal{N}_{\text{d}} 
+ \dfrac{\mathcal{D}^{\text{d}}_{\text{CSL}}}{2 m \gamma \hbar \omega}.
\label{eq:spp}
\end{align}
with $\sigma^{\infty}_{\mathcal{Z}_{\text{d}}\mathcal{P}_{\text{d}}}=0.$ The covariance matrix for the common mode takes a similar form, with $\omega_{\text{d}}$ and $\mathcal{D}^{\text{d}}_{\text{CSL}}$ replaced by $\omega$ and $\mathcal{D}^{\text{c}}_{\text{CSL}}$ respectively.

A repulsive Coulomb interaction, $\eta=1,$ leads to an increase in effective frequency $\omega_{\text{d}} = \omega\sqrt{1+\delta^{2}}$. This causes squeezing below the thermal level in $\sigma^{\infty}_{\mathcal{Z}_{\text{d}}\mathcal{Z}_{\text{d}}} = (\omega/2\omega_{\text{d}}) \mathcal{N}_{\text{d}} < \mathcal{N}/2$ in the absence of CSL, which persists even at high temperatures. The effect of collapse-induced noise is to provide \textit{additional} diffusive spread that can prevent such thermal state squeezing predicted by standard quantum theory. The detection of steady-state squeezing can thus be used to constrain the collapse rate parameter $\lambda_{\text{CSL}}$.

If collapse-induced diffusion exists, for $\eta = 1$, witnessing a steady-state squeezed thermal state with $\sigma^{\infty}_{\mathcal{Z}_{\text{d}}\mathcal{Z}_{\text{d}}} < \mathcal{N}/2$  implies (see Appendix~\ref{bound_squeezing}). 

\begin{align}\label{bound_thermal_sq}
    \mathcal{D}^{\text{d}}_{\text{CSL}}&<
    {\mathcal{N}m\gamma\hbar\omega}{\delta}^{2}\xrightarrow{{k_{\text{b}}}\text{T}\gg\hbar\omega}
    2{k_{\text{b}}}m\gamma \text{T}\delta^2.
\end{align}

This is our main result. Our proposal is to experimentally observe this squeezing, mediated by a \textit{weak} Coulomb interaction, which constrains collapse-induced diffusion by a small multiplier of $\delta^{2},$ tightening the constraints on collapse-induced diffusion. In principle, the smaller $\delta^{2}$ is, the better the constraint. In practice, experimentally detecting this squeezing requires experimental run-time $t_{\text{min}}\sim 1/(\delta^4 \gamma),$ where $\gamma$ is the mechanical damping coefficient. We provide some estimates for experimentally relevant parameters in Table~\ref{Table2}. 

A prominent advantage of our two-nanosphere set-up to constrain CSL over those with a single nanosphere is that the latter requires CSL diffusion to dominate over thermal dissipation to detect it~\cite{Nimrichter2}. In contrast, our proposal does not require discriminating CSL-induced diffusion from the thermal background, leading to a bound on  $\mathcal{D}^{\text{d}}_{\text{CSL}}$ as per Eq.~\eqref{bound_thermal_sq} by confirming $\sigma^{\infty}_{\mathcal{Z}_{\text{d}}\mathcal{Z}_{\text{d}}} < \mathcal{N}/2.$ 

Furthermore, if $\mathcal{N}$ is not fixed by the ambient gas (as is essential for low $T,\gamma$ regimes), its independent determination in the presence of the hypothetical, ever-present, $\mathcal{D}^{11}_{\text{CSL}}$ is impossible.  Using our two-particle setup with non-identical nanosphere, joint estimation of the position variance of each of the particles should enable simultaneous inference of $\mathcal{N} $ and $\mathcal{D}^{11}_{\text{CSL}}$ experimentally. This relies on $\mathcal{D}^{12}_{\text{CSL}}$ being negligible when $d \gg r_{\text{CSL}}.$

An attractive Coulomb interaction, $\eta=-1,$ leads to $\omega_{\text{d}} = \omega\sqrt{1-\delta^{2}}.$ In the absence of CSL, this leads to momentum-squeezing of the thermal state. 
However, such squeezing persists only at very low temperatures, $k_{b}\text{T}\sim \hbar{\omega_{\text{d}}}$. At higher temperatures, the momentum variance in Eq.~\eqref{eq:spp} is independent of $\omega_{\text{d}}.$

\textit{Upper Bounds on} $\lambda_{\text{CSL}}$:
To illustrate our proposal, we consider an experimental setup, as in Fig~\eqref{fig1}. Two charged nanospheres are simultaneously confined in a linear Paul trap and interact via the repulsive Coulomb potential. We seek to constrain $\lambda_{\text{CSL}}$ for $r_{\text{CSL}} \in [10^{-7},10^{-2}]$m using the parameters in Table~\ref{Table_1}. 

We find that stringent bounds on the CSL collapse rate can be obtained with micro-kelvin cooling of charged nanospheres. These bounds are computed using Eq.~\eqref{bound_thermal_sq} and the diffusion coefficients for macroscopic spheres, as detailed in Appendix~\ref{app:A} (see Eqs.~\eqref{Dcsl_11} and \eqref{CSL_correlating_diffusion}).

\begin{figure}[bp]
    \centering
    \includegraphics[width=\linewidth]{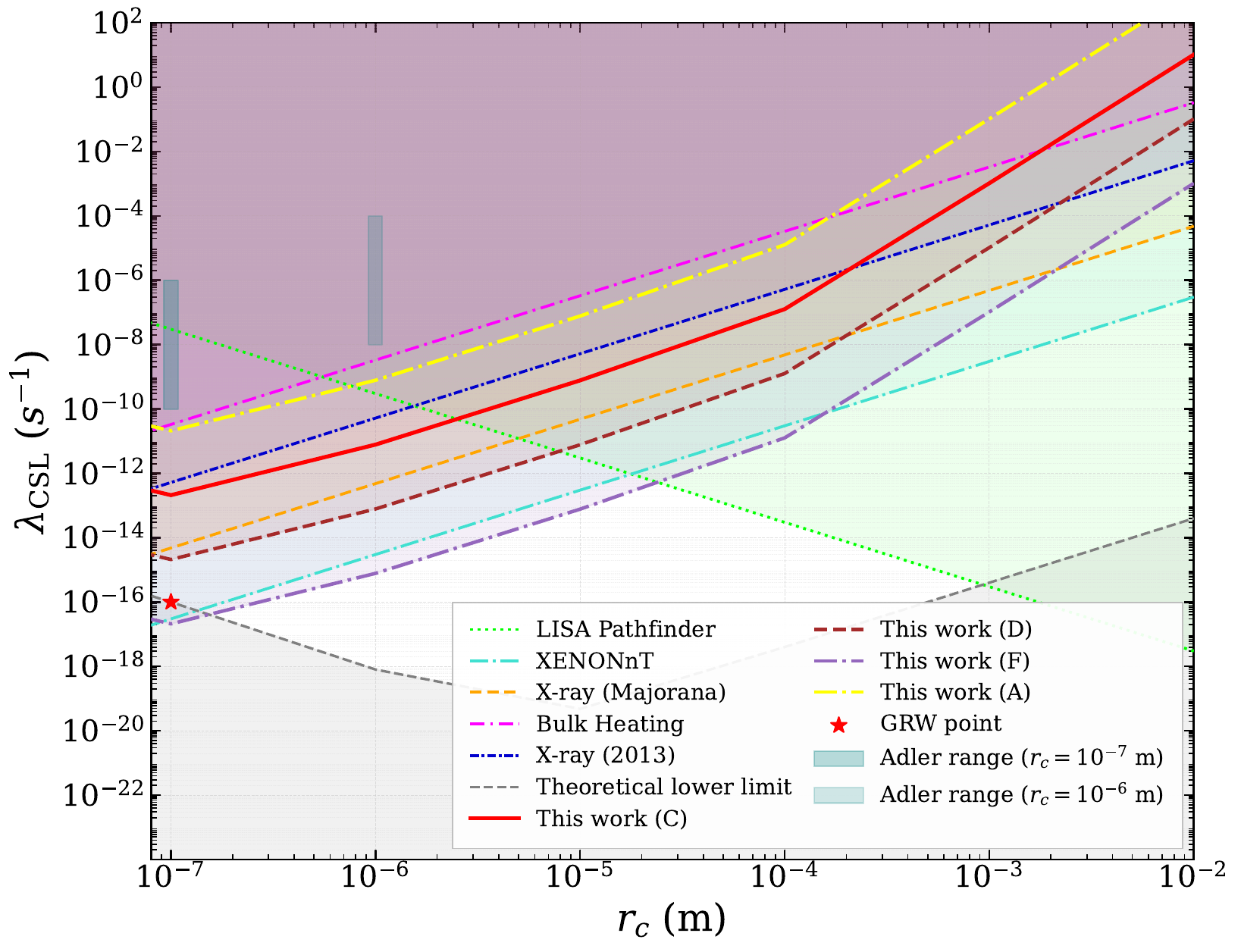}
    \caption{Our proposed bounds for parameters in Table 1 and different choices as in~\ref{Table2}. We include the current best bounds provided by XENONnT ~\cite{2jm3-4976}, and LISA Pathfinder~\cite{LISA,LISA2}. Previous bounds from X-ray experiments~\cite{Donadi2021,Majorana} and bulk-heating experiment~\cite{Adler,Alduino2017,Carlesso2022}.The grey-shaded region corresponds to the theoretically excluded region~\cite{Toro__2017}. The shaded vertical bars correspond to theoretical suggestions by Adler~\cite{Adler_bounds}, and the red dot to a suggestion by Ghirardi, Rimini, and Weber~\cite{GRW}.}
    \label{csl_exclusion_plot} 
\end{figure}

Our bounds are shown in Fig~\eqref{csl_exclusion_plot} for different choices of $T$ and $\gamma$ listed in Table~\ref{Table2}.
For choice C, and $r_{\text{CSL}} \lesssim 10^{-4}$m, our bound is better than the first X-ray emission experiment~\cite{Donadi2021} shown by the dotted blue line in Fig~\eqref{csl_exclusion_plot}.
At $r_{\text{CSL}}=10^{-7}\text{m},$ our bound constrains the collapse rate $\lambda_{\text{CSL}} < 2.09\times 10^{-14}\text{s}^{-1}$ which is an order of magnitude better than the first X-ray based bound of $\lambda_{\text{CSL}}<5.2\times 10^{-13}\text{s}^{-1}$ \cite{Donadi2021}. For choice F, our bounds can be more stringent than even those from the most recent X-ray emission experiment~\cite{2jm3-4976} (light-blue dot-dashed line).
Above $r_{\text{CSL}} \sim 10^{-4}\text{m},$ we see a sharp increase in our bound. This is due to effect of $\mathcal{D}^{12}_{\text{CSL}}$ which becomes significant around $r_{\text{CSL}}\sim d,$ reducing the effective diffusion coefficient of collapse noise acting on the differential mode. Thus, $\mathcal{D}^{\text{d}}_{\text{CSL}} = \mathcal{D}^{11}_{\text{CSL}}-\mathcal{D}^{12}_{\text{CSL}}$ becomes smaller, relaxing the constrain on $\lambda_{\text{CSL}}.$

\begin{table}[htbp]
\caption{Parameters used to constrain ${\lambda_{\text{CSL}}}$.
{Except temperature, the values are compatible with recent experiments ~\cite{Pontin2020,sympatheticcooling,ultrahighQ}.} 
}
\label{Table_1}
\begin{ruledtabular}
\begin{tabular}{lc}
\textbf{Parameter} & \textbf{Value} \\
\hline
Mass ($m$) & $0.44 \times 10^{-16}$~kg \\
Charge($q$) & $4\times 10^{-17}$~C\\
Radius ($R$) & 150~nm \\
Trap frequency ($\omega/{2\pi}$) & $500$~Hz \\
Mechanical damping rate ($\gamma$) & {$10^{-4}$}~s$^{-1}$ \\
Temperature ($T$) & 10~$\mu$K \\
Mean separation ($d$) & $1 \times 10^{-4}$~m \\
Frequency shift ($\delta^{2}$) & $0.13$ \\
\end{tabular}
\end{ruledtabular}
\end{table}

Bounds from X-ray emission experiment, however, become completely irrelevant when a coloured noise spectrum for the collapse models is considered. It is physically reasonable to associate the collapse mechanism with the action of an underlying field, potentially of cosmological origin. This is a suggestive physical mechanism for the generation of weak correlations between two \textit{non-interacting particles} via $\mathcal{D}^{12}_{\text{CSL}}$ in Eq.\eqref{eq:ME}. Noise spectra associated with such a physical field are never white, as they exhibit finite-time correlations. It is therefore more realistic to consider a colored noise spectrum for the collapse models. Such models have been developed \cite{Adler_2007} and used to constrain the parameters of the collapse model \cite{Carlesso_2018,Toro__2017}. These considered an exponentially decaying noise autocorrelation with a frequency cut-off $\Omega.$ The effect of collapse noise at frequencies above $\Omega$ is highly suppressed. 

We show in Appendix~\ref{app:B} that the bounds on $\lambda_{\text{CSL}}$ obtained using our set-up are robust against such modifications, whereas, for $\Omega \sim 10^{12}$ Hz \cite{Bassi_2010,Carlesso_2018}, the bounds obtained from X-ray-based experiments become severely weakened. Conservatively, the bound from the XENONnT collaboration~\cite{2jm3-4976} becomes $\lambda_{\text{CSL}} < 7.09\times 10^{-3}\text{s}^{-1}$ at $r_{\text{CSL}} = 10^{-7}\text{m}$ (Appendix~\eqref{Xray-cCSL}).Moreover, for $\Omega \sim 10^{11}\text{Hz}$, even bulk-heating experiments do not appreciably constrain collapse models~\cite{Carlesso_2018}. Only low-frequency mechanical experiments remain robust against such modifications and provide the most reliable bounds. Being of this nature, our work can provide the most stringent bounds on the CSL model.

\begin{table}[h!]
\centering
\caption{Bounds on $\lambda_{\text{CSL}}$ {at $r_{\text{CSL}} =10^{-7}\text{m}$}  and corresponding experimental run-times (in days) for different {choices of } $T$ and $\gamma.$ All other parameters same as in Table~\ref{Table_1} including $\delta^2=0.13.$
}
\label{Table2}
\begin{ruledtabular}
\begin{tabular}{ccccc}
{Choice} & $T$ & $\gamma$ (s$^{-1}$) & $t_{\min}$ ({days}) & $\lambda_{\text{CSL}}$(s$^{-1}$) \\
\hline
A & $100~\mathrm{mK}$  & $1.0\times 10^{-4}$ & ${14}$  & $2.09\times 10^{-11}$ \\
B & $1~\mathrm{mK}$    & $1.0\times 10^{-3}$ & ${1.4}$ & $2.09\times 10^{-12}$ \\
C & $1~\mathrm{m K}$   & $1.0\times 10^{-4}$ & $14$    & $2.09\times 10^{-13}$ \\
D & $10~\mathrm{\mu K}$ & $1.0\times 10^{-4}$ & $14$    & $2.09\times 10^{-15}$ \\
E & $10~\mathrm{\mu K}$ & $1.0\times 10^{-5}$ & $140$   & $2.09\times 10^{-16}$ \\
F & $1~\mathrm{\mu K}$  & $1.0\times 10^{-5}$ & $140$   & $2.09\times 10^{-17}$ \\
\end{tabular}
\end{ruledtabular}
\label{tab:lambdaCSL}
\end{table}

\textit{Dynamics at short times}\label{shortime_squeezing}-
Preceding bounds correspond to the long-time regime $t\gg 1/\gamma.$ We next explore the short times, $t \ll 1/\omega $, dynamics driven by white-noise collapse models. The squeezing of the differential mode is characterized by the minimum eigenvalue, $\nu(t)$, of the covariance matrix $\sigma_{\text{d}}(t)$ of this mode. Up to $\mathcal{O}(t)$
\begin{equation}
        \nu(t) = \frac{1}{2}\left(\mathcal{N} + \frac{t}{\hbar m \omega}\left(\mathcal{D}_{\text{eff}} - \sqrt{\mathcal{D}^{2}_{\text{eff}} + \delta^{4}\omega^{4}\hbar^{2}m^{2}\mathcal{N}^{2}}\right)\right)
\end{equation}
where $\mathcal{D}_{\text{eff}} = \mathcal{D}^{\text{d}}_{\text{CSL}} + \mathcal{D}_{\text{th}} - m\hbar\omega\gamma\mathcal{N}.$ This shows that the system features a squeezed thermal state because $\nu(t) < \mathcal{N}/2$. Incidentally, if the two particles are initially cooled to their ground state, then for $t \ll 1/\omega,$ $\nu(t) < 1/2$, indicating that quantum squeezing is robust against both dissipative and diffusive noise at short times. Remarkably, at high temperatures, i.e, ${k_{\text{b}}}T\gg\hbar\omega$, $\mathcal{D_{\text{eff}}} \rightarrow \mathcal{D_{\text{CSL}}}$.
Thus, only the contribution of the collapse-diffusive noise remains.

At short times, a weak Coulomb-mediated entanglement test can be used to strongly constrain $\lambda_{\text{CSL}}$. For two quantum oscillators, initialized in their ground state, evolving under mutual Coulomb coupling in the presence of diffusive and dissipative noise, the criteria for entanglement at short times are given by {(see Eq.~\eqref{entanglement_bound} in Appendix \eqref{app:C})}
\begin{align}
     4\left(\mathcal{D}^{11}_{\text{CSL}} + \mathcal{D}_{\text{th}}\right)^{2} < 4(\mathcal{D}^{12}_{\text{CSL}})^{2} + \delta^{4}m^{2}\omega^{4}\hbar^{2}.
\end{align}
For $d \gg r_{\text{c}}$, a conservative entanglement-based bound on the collapse-induced diffusion becomes $\mathcal{D}^{11}_{\text{CSL}} < m\omega^{2}\hbar\delta^{2}/2$ (see Appendix~\eqref{app:C}). 

Observation of Coulomb-mediated entanglement, characterized by $\delta^{2} = 1.4 \times 10^{-3}$, between two charged nanospheres initialized in ground state, at an environmental temperature of $1\text{K}$ requires a mechanical quality factor of $Q = 10^{11}.$ Values of $Q \sim 10^{10}$ have already been achieved~\cite{ultrahighQ}. In this regime, detecting entanglement at short times constrains the collapse rate to $\lambda_{\text{CSL}} < 1.7 \times 10^{-12}~\text{s}^{-1}$, and more ambitiously, constraining $\lambda_{\text{CSL}} < 4\times 10^{-14} \text{s}^{-1}$ might also be possible. The full exclusion plot is presented in {Fig~\eqref{fig:entanglement_bound}} in Appendix~\eqref{app:C}. This bound can be further improved by reducing $\delta^{2}$. Such bounds require initial ground-state cooling and $Q\sim 10^{11}$ for $\omega/(2\pi)\sim 1$~kHz. This is feasible using feedback cooling strategies in levitated charged nano-particle experiments~\cite{Dania2022}.

\textit{Discussion} - 
We have shown that experiments with two cooled, charged nanospheres can provide some of the most stringent bounds on CSL models -- both white and coloured. 
The prospects of probing the DP model using our scheme is addressed in detail in Appendix~\eqref{app:D}. To obtain a meaningful bound on the parameter $R_{0}$ of the DP model requires satisfying  $G \rho\sqrt{\pi} > 9\delta^{2}\omega\gamma \mathcal{N}.$ This requires $\mathcal{N}\sim \mathcal{O}(1)$, very low damping and extremely low $\delta^{2}\sim 10^{-7}$, which in turn implies extremely large experimental run-time.
Probing the DP model with Coulomb-mediated entanglement requires satisfying similar conditions, which require testing for extremely weak entanglement (see Appendix~\eqref{dp_entanglement_bound}).
Probing the DP model with our scheme thus remains challenging.

One of us (AD) previously proposed testing the Schr\"odinger-Newton (SN) model of semi-classical gravity~\cite{Bahrami_2014_SN} by comparing the relative quantum squeezing of the differential mode to that of the common mode mediated by Newtonian potential~\cite{Datta_2021}. While it focused on observing quantum squeezing in the steady state, witnessing a squeezed thermal state can serve the same purpose. As we have shown that both squeezing in thermal state and quantum squeezing (for oscillators initially cooled to their ground state) are found to be robust against noise, the experimental scheme proposed in~\cite{Datta_2021} could benefit from targeting squeezing at short times.

\section*{Acknowledgments}

AD acknowledges the UK STFC
“Quantum Technologies for Fundamental Physics” programme
(Grant Numbers ST/T006404/1, ST/W006308/1
and ST/Y004493/1), and PB acknowledges the UK EPSRC and STFC (Grant Numbers EP/W029626/1 and ST/W006170/1)  for support.

\bibliography{bibliography}

\onecolumngrid
\clearpage
\appendix

\section{Collapse Dynamics With Two Massive Systems.}
\label{app:A}

In this section, we provide the derivation of the linear diffusion coefficients induced by the Collapse models. We first start with the Diosi-Penrose model, the continuous spontaneous localization model is very similar, differing only in the choice of noise kernel. 

\subsection{Diosi-Penrose Collapse}\label{DP_Eq}
The evolution of quantum systems with Diosi-Penrose (DP) collapse is described by:

\begin{align}
    \frac{\partial\hat{\rho}}{\partial t} &= -\frac{i}{\hbar}[\hat{H},\hat{\rho}(t)] + \mathcal{L}[\hat{\rho}(t)],
\end{align}

Where, $\hat{H}$ is system Hamiltonian, and $\mathcal{L}[\hat{\rho}(t)]$ reads\cite{Role_of_gravity,DIOSI1987377}:

\begin{align}
    \mathcal{L}[\hat{\rho}] = -\frac{G}{2\hbar}\int\int\frac{d^{3}\mathbf{x}d^{3}\mathbf{y}}{|\mathbf{x}-\mathbf{y}|} [\hat{M}(\mathbf{x}),[\hat{M}(\mathbf{y}),\hat{\rho}]],
\end{align}

\begin{equation}
\hat{M}(\mathbf{x}) =\sum_{\alpha \in \{1,2\}} \sum_{i=1}^{N}  
\frac{m^{(\alpha)}_i}{(2\pi R_0^2)^{3/2}} 
\exp\left( -\frac{|\mathbf{\hat{r}}^{(\alpha)}_i - \mathbf{x}|^2}{2 R_0^2} \right),
\end{equation}

Here, $R_0$ is the DP parameter which sets the smallest length scale in the model. For a rigid body, we can write, for each i-th particle ($i=(1,2,.., N)$): $ \hat{\mbf r}^{(\alpha)}_{i} = \hat{\mbf r}^{(\alpha)}_{\text{CM}} +  {\mbf r}^{(\alpha)}_{i,\text{rel}}$, where, $\hat{\mbf r}^{(\alpha)}_{\text{CM}}$ is the center of mass (COM) position operator of the massive system indexed $\alpha$ and $\mbf r^{(\alpha)}_{i,\text{rel}}$ are the classical relative coordinates.

Writing $\hat{M}(\mbf x)$ in the Fourier space, and in terms of the COM position operator, we can isolate the center of mass contribution from the relative coordinates. We can write:

\begin{align}
    \hat{M}(\mbf x) &= \sum_{\alpha \in \{1,2\}} \sum_{i=1}^{N}  
\frac{m^{(\alpha)}_i}{(2\pi)^{3}} 
\int \text{d}^{3}\mbf q\exp\left( -i \mbf q.(\hat{\mbf r}^{(\alpha)}_{\text{CM}} + {\mbf r}^{(\alpha)}_{i,\text{rel}} -{\mbf x}) \right)e^{-R^{2}_{0}q^{2}/2},\\
&=\sum_{\alpha \in \{1,2\}}\int \text{d}^{3}\mbf q \frac{1}{(2\pi)^{3}} 
 \underbrace{\sum_{i=1}^{N} m^{(\alpha)}_i e^{-i\mbf q. {\mbf r}^{(\alpha)}_{i,\text{rel}}}}_{{{{\tilde{\mu}}}^{(\alpha)}(\mbf q)}}
\exp\left( -i \mbf q.(\hat{\mbf r}^{(\alpha)}_{\text{CM}}   -{\mbf x}) \right)e^{-R^{2}_{0}q^{2}/2},\\
&=\sum_{\alpha \in \{1,2\}}\int \text{d}^{3}\mbf q   
\frac{1}{(2\pi)^{3}} {{\tilde\mu}^{(\alpha)}}(\mbf q)
\exp\left( -i \mbf q.(\hat{\mbf r}^{(\alpha)}_{\text{CM}}   -{\mbf x}) \right)e^{-R^{2}_{0}q^{2}/2}.
\end{align}

Where we have defined $\tilde{\mu}^{(\alpha)}(\mbf q) = \sum^{N}_{i=1}m^{(\alpha)}_{i}e^{-i\mbf q.{\mbf r}^{(\alpha)}_{i,\text{rel}}}$ as the Fourier transform of the mass density. We also make use of the identity:

\begin{align}
  \frac{1}{|\mbf x-\mbf y|} = \frac{1}{2\pi^{2}}\int \text{d}^{3}\mbf p\frac{e^{-i\mbf p.(\mbf x -\mbf y)}}{p^{2}}.  
\end{align}

Putting back everything together, we can then obtain a simplified expression for $\mathcal{L}[\hat{\rho}(t)]:$

\begin{align}
    \mathcal{L}[\hat{\rho}(t)] 
    &= -\frac{G}{2\hbar} \int \! \text{d}^3 \mathbf{x} \, \text{d}^3 \mathbf{y} \frac{1}{2\pi^{2}}\frac{1}{(2\pi)^{6}}
    \sum_{\alpha,\beta \in \{1,2\}} 
    \int \! \text{d}^3 \mathbf{p} \, \text{d}^3 \mathbf{q} \, \text{d}^3 \mathbf{k} \,
    \frac{1}{p^2} {\tilde{{\mu}}^{(\alpha)}}(\mathbf{q}) \tilde{{\mu}}^{(\beta)}(\mathbf{k}) \notag \\
    &\quad \times e^{-\frac{R^{2}_{0}(q^{2}+k^{2})}{2}}e^{-i \mathbf{p} \cdot ({\mathbf{x}} - {\mathbf{y}})} 
    e^{i \mathbf{q} \cdot {\mathbf{x}}} 
    e^{i \mathbf{k} \cdot {\mathbf{y}}} 
    \left[ e^{i \mathbf{q} \cdot \hat{\mathbf{r}}^{(\alpha)}_{\text{CM}}}, 
    \left[ e^{i \mathbf{k} \cdot \hat{\mathbf{r}}^{(\beta)}_{\text{CM}}}, \hat{\rho}(t) \right] \right].
\end{align}

Evaluating some of the integrals straightforwardly leads to the following:
\begin{align}
    \mathcal{L}[\hat{\rho}(t)] 
    &= -\frac{G}{2\hbar} \cdot \frac{1}{2\pi^{2}} 
    \sum_{\alpha,\beta \in \{1,2\}} \int \! \text{d}^{3} \mathbf{p} \, 
    \frac{1}{p^2} \, {\tilde{{\mu}}^{(\alpha)}}(\mathbf{p}) \, {\tilde{{\mu}}^{(\beta)}}(-\mathbf{p}) \, 
    e^{-R^{2}_{0} p^{2}} \notag \\
    &\quad \times 
    \left[ e^{i \mathbf{p} \cdot \hat{\mathbf{r}}^{(\alpha)}_{\text{CM}}}, 
    \left[ e^{-i \mathbf{p} \cdot \hat{\mathbf{r}}^{(\beta)}_{\text{CM}}}, \hat{\rho}(t) \right] \right].
\end{align}

Expanding $e^{i\mbf p\cdot  \hat{\mbf r}_{\text{CM}}}$ and retaining only the first order terms and projecting oscillations along the $z$-axis, we obtain:

\begin{align}\label{lin_meq}
    \mathcal{L}[\hat{\rho}(t)] 
    &= -\frac{G}{4\hbar\pi^{2}} 
    \sum_{\alpha,\beta \in \{1,2\}} \int \! \text{d}^{3} \mathbf{p} \, 
    \frac{p_{z}^{2}}{p^2} \, {\tilde{{\mu}}^{(\alpha)}}(\mathbf{p}) \, {\tilde{{\mu}}^{(\beta)}}(-\mathbf{p}) \, 
    e^{-R^{2}_{0} p^{2}} \notag \\
    &\quad \times 
    \left[ \hat{z}^{(\alpha)}_{\text{CM}}, 
    \left[ \hat{z}^{(\beta)}_{\text{CM}}, \hat{\rho}(t) \right] \right].
\end{align}

Comparing with $\mathcal{L}[\hat{\rho}(t)] = -\frac{1}{\hbar^{2}}\sum_{\alpha,\beta\in\{1,2\}}\mathcal{D}^{\alpha\beta}_{\text{DP}}[\hat{z}_{\alpha},[\hat{z}_{\beta},\hat{\rho}(t)]$, we obtain:

\begin{align}
    \mathcal{D}^{11}_{\text{DP}} =  \mathcal{D}^{22}_{\text{DP}} &= \frac{G\hbar}{4\pi^{2}}\int \! \text{d}^{3} \mathbf{p} \, 
    \frac{p_{z}^{2}}{p^2} \, |\tilde{\mu}^{11}(\mathbf{p})|^{2} 
    e^{-R^{2}_{0} p^{2}},\\
    \mathcal{D}^{12}_{\text{DP}} &= \frac{G\hbar}{4\pi^{2}}\int \! \text{d}^{3} \mathbf{p} \, 
    \frac{p_{z}^{2}}{p^2} \, \tilde{\mu}^{11}(\mathbf{p})\tilde{\mu}^{22}(-\mathbf{p})
    e^{-R^{2}_{0} p^{2}}.
\end{align}

For many particles within a volume \( R_0^3 \), we can replace the discrete model with a continuum model, and define \cite{Bassi_2003,Nimrichter2}:
\[
\tilde{\mu}(\mathbf{p}) = \int \! \mathrm{d}^3 \mathbf{r} \, \rho(\mathbf{r}) \, e^{-i \mathbf{p} \cdot \mathbf{r}},
\]
where $\rho(\mbf r)$ is the mass-density.

For a homogenous sphere of radius $R$ and mass $m$, approximating as a Gaussian mass density~\cite{Role_of_gravity}, i.e, $\mu(r) = \frac{m}{(2\pi R^{2})^{3/2}}e^{-\frac{r^{2}}{2 R^{2}}}$ for simplicity, we obtain:

\begin{align}\label{DP_coeff_appendix}
    \mathcal{D}_{\text{DP}}^{11} =\mathcal{D}_{\text{DP}}^{22} &=\frac{G\hbar m^{2}}{12\sqrt{\pi} \mathcal{R}_{D}^{3}}\\
    \mathcal{D}_{\text{DP}}^{12} &= \frac{G\hbar m^{2}}{4\sqrt{\pi}\mathcal{R}_{D}^{3}}\left(4\frac{\mathcal{R}_{D}^{2}}{d^{2}} + 1\right)e^{-\frac{4d^{2}}{\mathcal{R}_{D}^{2}}} - \frac{G\hbar m^{2}}{d^{3}}\text{Erf}\left(\frac{d}{2\mathcal{R}_{D}}\right).
\end{align}

Where, $\mathcal{R}_{D} = \sqrt{R^{2}+R^{2}_{0}}$ serves as the effective divergence regulator.

\subsection{{Continuous} Spontaneous Localization (CSL) Model}\label{CSL_eq}

The CSL master equation reads \cite{Bassi_2003,Nimrichter2}:

\begin{align}
    \frac{\partial\hat{\rho}(t)}{\partial t} = -\frac{i}{\hbar}[\hat{H},\hat{\rho}(t)] -\frac{\lambda_{\text{CSL}} (4\pi r^{2}_{\text{CSL}})^{3/2}}{2m^{2}_{0}}\int d^{3}\mbf x[\hat{M}(\mbf x),[\hat{M}(\mbf x),\hat{\rho}(t)]].
\end{align}

With:

\begin{equation}
\hat{M}(\mathbf{x}) =\sum_{\alpha \in \{1,2\}} \sum_{i=1}^{N}  
\frac{m^{(\alpha)}_i}{(2\pi r_\text{CSL}^2)^{3/2}} 
\exp\left( -\frac{|\mathbf{r}^{(\alpha)}_i - \mathbf{x}|^2}{2 r_{\text{CSL}}^2} \right),
\end{equation}

The DP master equation and the CSL master equation has very similar structures, here the collapse width $r_{\text{CSL}}$ plays the role of $R_{0}$ and we have the kernel $\delta(\mbf x -\mbf y)$ instead of $\frac{1}{|\mbf x - \mbf y|}.$ Following the same lines of computations as in the previous section, we arrive at:

\begin{align}\label{lin_csl}
    \mathcal{L}[\hat{\rho}(t)] 
    &=\frac{-\lambda_{\text{CSL}}r^{3}_{\text{CSL}}}{2 m^{2}_{0}\pi^{3/2}}
    \sum_{\alpha,\beta \in \{1,2\}} \int \! \text{d}^{3} \mathbf{p} \, 
    {p_{z}^{2}} \, \tilde{\mu}^{(\alpha)}(\mathbf{p}) \, \tilde{\mu}^{(\beta)}(-\mathbf{p}) \, 
    e^{-r^{2}_{\text{CSL}} p^{2}} \notag \\
    &\quad \times 
    \left[ \hat{z}^{(\alpha)}_{\text{CM}}, 
    \left[ \hat{z}^{(\beta)}_{\text{CM}}, \hat{\rho}(t) \right] \right].
\end{align}

Again, comparing with $\mathcal{L}[\hat{\rho}(t)] = -\frac{1}{\hbar^{2}}\sum_{\alpha,\beta\in\{1,2\}}\mathcal{D}^{\alpha\beta}_{\text{CSL}}[\hat{z}_{\alpha},[\hat{z}_{\beta},\hat{\rho}(t)]$, we obtain:

\begin{align}
    \mathcal{D}^{11}_{\text{CSL}} =  \mathcal{D}^{22}_{\text{CSL}} &= \frac{\lambda_{\text{CSL}}\hbar^{2} r^{3}_{\text{CSL}}}{2 m^{2}_{0}\pi^{3/2}}\int \! \text{d}^{3} \mathbf{p} \, 
    {p_{z}^{2}} \, |\tilde{\mu}^{11}(\mathbf{p})|^{2} 
    e^{-r^{2}_{\text{CSL}} p^{2}},\\
    \mathcal{D}^{12}_{\text{CSL}} &= \frac{\lambda_{\text{CSL}}\hbar^{2}r^{3}_{\text{CSL}}}{2 m^{2}_{0}\pi^{3/2}}\int \! \text{d}^{3} \mathbf{p} \, 
    {p_{z}^{2}} \, \tilde{\mu}^{11}(\mathbf{p})\tilde{\mu}^{22}(-\mathbf{p})
    e^{-r^{2}_{\text{CSL}} p^{2}}.
\end{align}

For a rigid homogeneous sphere of constant density $\mu(r) = \frac{m}{V}$; we obtain:

\begin{align}\label{Dcsl_11}
     \mathcal{D}^{11}_{\text{CSL}} =\mathcal{D}^{22}_{\text{CSL}} &= \lambda_{\text{CSL}}\frac{\hbar}{r^{2}_{\text{CSL}}}\left(\frac{3m^{2}}{ m^{2}_{0}}\frac{r^{6}_{\text{CSL}}}{R^{6}}\left(\left(1+\frac{R^{2}}{2 r^{2}_{\text{CSL}}}\right)e^{\frac{-R^{2}}{r^{2}_{\text{CSL}}}} + \left(-1 + \frac{R^{2}}{2r_{\text{CSL}}^{2}}\right)\right)\right).
\end{align}

It is difficult to obtain an analytic formula for $\mathcal{D}^{12}_{\text{CSL}}$ for a homogeneous rigid body.
We approximate it, assuming a Gaussian mass distribution $\mu(r) = \frac{m}{(2\pi R^{2})^{3/2}}e^{-\frac{r^{2}}{2 R^{2}}},$ obtaining
\begin{align}\label{CSL_correlating_diffusion}
 \mathcal{D}^{12}_{\text{CSL}} &= \frac{\lambda_{\text{CSL}}}{8}\left(\frac{\hbar}{r_{\text{CSL}}}\right)^{2}\left(\frac{m}{m_{0}}\right)^{2}r^{5}_{\text{CSL}}\left(\frac{2}{\mathcal{R}^{5}_{\text{CSL}}}-\frac{d^{2}}{\mathcal{R}^{7}_{\text{CSL}}}\right)e^{-\frac{d^{2}}{4\mathcal{R}^{2}_{\text{CSL}}}}.
 \end{align}
Here, we have defined, $\mathcal{R}_{\text{CSL}} = \sqrt{R^{2}+r_{\text{CSL}}^{2}}.$

\section{Bounds on Collapse Models from Witnessing Steady-State Thermal Variance Reduction}\label{bound_squeezing}

In this section, we prove the bound presented in {Eq.~\eqref{bound_thermal_sq}}. As discussed in the main text, we propose to witness a reduction in the thermal variance induced by the frequency shift factor, $\delta^2$. Collapse-induced diffusion acts to suppress this effect, the occurrence of which can then constrain the noise of the collapse. 

The signal is the observation: $\sigma_{\mathcal{Z}_{\text{d}}\mathcal{Z}_{\text{d}}} < \frac{\mathcal{N}}{2}.$ Using the expressions in {Eq.~\eqref{cov_matrix}, we find}
\begin{align}
    \mathcal{D}^{\text{d}}_{\text{CSL}} < \frac{m\gamma\hbar\omega^{2}_{\text{d}}\mathcal{N}}{\omega}\left(1-\frac{\omega}{\omega_{\text{d}}}\frac{\mathcal{N}_{\text{d}}}{\mathcal{N}}\right).
\end{align}
Written in the form above, it is not very illuminating. We, however, note the following inequality satisfied by the $\coth(x)$ function:

\begin{align}
    \frac{1}{\beta}\leq\frac{\coth(\beta x)}{\coth(x)}\le 1, \hspace{0.5cm}\text{For $\beta >1.$}
\end{align}

\textit{Proof:}
Observe that $\coth(x)$ is a monotonically decreasing function, therefore, trivially, $\coth(\beta x) < \coth(x)$ for $\beta >1,$ which proves the upper bound.
To prove the lower bound, we observe that the function $g(x) = x\coth(x)$ is a strictly increasing function. This can be verified by standard methods and we omit the proof here. We then have $g(\beta x)>g(x)$ for $\beta>1,$ and the lower bound trivially follows.

Using the above proved inequality, we have $\frac{\omega}{\omega_{\text{d}}}\frac{\mathcal{N}_{\text{d}}}{\mathcal{N}}\geq \frac{\omega^2}{\omega^2_{\text{d}}}.$ After some simple algebra, we finally obtain:

\begin{align}
    \mathcal{D}^{\text{d}}_{\text{CSL}} < \frac{m\gamma\hbar\omega^{2}_{\text{d}}\mathcal{N}}{\omega}\left(1-\frac{\omega}{\omega_{d}}\frac{\mathcal{N}_{d}}{\mathcal{N}}\right) < \mathcal{N}m\gamma\hbar\omega\delta^2 \xrightarrow{k_{\text{b}}\text{T}\gg \hbar\omega} 2mk_{\text{b}}\text{T}\gamma\delta^2.
\end{align}

\section{Coloured Collapse Models}\label{app:B}

Coloured collapse models consider the prospects of the mechanism of wave-function collapse being induced by a physical, classical stochastic field. As noise spectra associated with physical fields are never white, a more realistic way to model the action of collapse noise is to consider a non-trivial correlation function. Coloured collapse models have previously been considered to bound the parameters of continuous spontaneous localization model \cite{Toro__2017, Carlesso_2018}. In these works, an exponentially decaying autocorrelation has been adopted primarily for mathematical convenience. We point out that in classical stochastic theory, one finds that the only non-trivial continuous stationary Gaussian Markov process is the Ornstein–Uhlenbeck process, which has an exponential autocorrelation~\cite{2deae53e-bc5e-3947-9dd3-8ae0b7e9d718}. Hence, under the assumption that the collapse noise is Markovian, the autocorrelation function is uniquely determined to be $\langle \zeta(t)\zeta(t')\rangle \sim f(t-t')$, where $f(t-t') =\frac{\Omega e^{-\Omega|t-t'|}}{2}.$ For a non-Markovian noise process, no such unique correlation choice exists. In this model, the coloured CSL (cCSL) noise has a correlation time of $\mathcal{O}(\frac{1}{\Omega}).$ In the limit $\Omega \rightarrow \infty$ we recover the white noise model. 

Working through the same set-up for the dynamics of the differential mode considered in the main-text, we find that the collapse-induced diffusion contributes the following terms to the steady-state covariance matrix of the differential mode:
\begin{align}
    \sigma^{\infty}_{\mathcal{Z}_{\text{d}}\mathcal{Z}_{\text{d}}} &= \frac{\omega}{2\omega_{\text{d}}} \mathcal{N}_{\text{d}} +  \frac{\mathcal{D}^{\text{d}}_{\text{CSL}}\omega}{ 2m\gamma\hbar\omega^{2}_{\text{d}}}\left(\frac{\Omega(\Omega+\gamma)}{ \Omega^{2} + \omega^{2}_{\text{d}} +\frac{\Omega\gamma}{2}}\right),\\
    \sigma^{\infty}_{\mathcal{Z}_{\text{d}}\mathcal{P}_{\text{d}}} &=0,\\
    \sigma^{\infty}_{\mathcal{P}_{\text{d}}\mathcal{P}_{\text{d}}} &= \frac{\omega_{\text{d}}}{2\omega} \mathcal{N}_{\text{d}}+\frac{\mathcal{D}^{\text{d}}_{\text{CSL}}}{2 m\gamma \hbar\omega}\left(\frac{\Omega(\Omega+\gamma)}{ \Omega^{2} + \omega^{2}_{\text{d}} +\frac{\Omega\gamma}{2}}\right).
\end{align}

In the $\gamma \ll 1$ regime, we essentially have an effective frequency-dependent collapse diffusion coefficient $\mathcal{D}_{\text{CSL}} (\omega_{\text{d}},\Omega) =\mathcal{D}_{\text{CSL} }\left(\frac{\Omega^{2}}{\Omega^{2} + \omega^{2}_{\text{d}}}\right)$. In practice, this means that the finite noise correlation time cuts off the collapse action at frequencies above $\Omega$. We have three regimes of interest, $\Omega \gg \omega_{\text{d}}$, $\Omega \approx\omega_{\text{d}}$, and $\Omega \ll \omega_{\text{d}}$. It is clear that the constraints on the collapse parameters remain robust for $\Omega \gg \omega_{\text{d}}$ and relax by a factor of 2 for $\Omega \approx \omega_{\text{d}},$ whereas the constraints become much less severe for $\Omega \ll \omega_{\text{d}}.$ 

The universality of the collapse noise motivates the assumption of a cosmological origin of cCSL. Such considerations suggest $\Omega \sim 10^{12}$ Hz \cite{Bassi_2010,Carlesso_2018}, in this regime, the bounds from X-ray radiation search becomes severely weakened (see:Appendix~\eqref{Xray-cCSL}), and the strongest robust bounds on collapse models comes from bulk-heating experiments~\cite{Adler}, and as argued in \cite{Carlesso_2018} for $\Omega \sim 10^{11}\text{Hz}$ even bulk-heating experiments do not appreciably constrain collapse models. Therefore, only low-frequency mechanical experiments remain robust against such modifications and provide the most reliable bounds.

Since, in our case, $\omega \ll \Omega,$ the coloured-collapse models reduce to the white-noise models, and the same conclusions apply to the entanglement test discussed in Appendix~\eqref{app:C}, i.e., the bounds obtained via a short-time Coulomb entanglement test between two ground-state quantum oscillators remain unaffected by modifications in the spectrum.

\subsection{X-ray Emission With Colored Collapse Models}\label{Xray-cCSL}

The diffusive motion induced by spontaneous collapse models leads to random acceleration of atoms and thus emission of Electromagnetic radiation from its charged constituents. Lack of observation of such predicted radiation can be used to bound collapse model parameters.

The expression for radiation emission in white noise collapse models, under the approximation of coherent proton and independent electron radiation emissions, is given by \cite{Donadi2021}:

\begin{equation}
    \frac{d\Gamma^{\text{CSL}}}{dE} = \frac{\hbar e^{2}\lambda_{\text{CSL}}\left(N^{2}_{p}+N_{e}\right)}{4\pi^2\epsilon_{0}c^{3}r^{2}_{\text{CSL}}m^{2}_{0}E}.
\end{equation}

Here, $e$ denotes electronic charge, $m_0$ the mass of a nucleon, $N_{p}$ and $N_{e}$ are the number of protons and electrons, respectively, and $E$ corresponds to the energy of the spontaneously emitted photon. The expression for a more general situation can be found in \cite{K.Piscicchia}. 

If we introduce a frequency cut-off into the spectrum of collapse noise, $\Omega$, the expression gets modified to \cite{K.Piscicchia,Carlesso_2018}:

\begin{equation}\label{cCSL_x_ray}
     \frac{d\Gamma^{\text{CSL},\Omega}}{dE} =  \frac{d\Gamma^{\text{CSL}}}{dE}\left(\frac{E^{2}_{\Omega}}{E^{2}_{\Omega}+E^{2}}\right).
\end{equation}

Where $E_{\Omega} = \hbar\Omega.$ The XENONnT collaboration \cite{2jm3-4976} considered X-ray emission in the energy range, $E \in [1-140]\text{KeV},$ and constrained collapse parameters to: $\frac{\lambda_{\text{CSL}}}{r^{2}_{\text{CSL}}} < 3 \times 10^{-3}\text{s}^{-1}\text{m}^{-2}$, the analysis considered radiation in the more general case discussed in \cite{K.Piscicchia}, however, it is immaterial for our discussion. The key observation is that we can consider the effect of cCSL to introduce an effective parameter, $\lambda_{\text{cCSL}} = \lambda_{\text{CSL}}\left(\frac{E^{2}_{\Omega}}{E^{2}_{\Omega}+E^{2}}\right).$ For cCSL, this constraint translates to:

\begin{align}
    \lambda_{\text{cCSL}} < 3\times 10^{-3}\text{s}^{-1}\text{m}^{-2}r^{2}_{\text{CSL}}\left(\frac{E^{2}_{\Omega}+E^{2}}{E^{2}_{\Omega}}\right).
\end{align}

Considering $\Omega \sim 10^{12}\text{Hz}$ \cite{Carlesso_2018, Bassi_2003}, we have $E_{\Omega} \sim 6.5\times 10^{-4} \text{eV} \ll 1\text{KeV} \sim E,$ as a result, the bound on $\lambda_{\text{CSL}}$ is severely weakened, and conservatively we have: $\frac{\lambda_{\text{CSL}}}{r^{2}_{\text{CSL}}} < 7.09 \times 10^{11}\text{s}^{-1}\text{m}^{-2}.$

Similarly, the spontaneous radiation, according to a Markovian DP model is given by~\cite{K.Piscicchia}:

\begin{align}
    \frac{d\Gamma^{\text{DP}}}{dE} = \frac{Ge^{2}\left(N^{2}_{p}+N_{e}\right)}{12\pi^{5/2}\epsilon_{0}c^{3}R^{3}_{0}E}
\end{align}

A coloured spectrum, leads to a similar modification as~\eqref{cCSL_x_ray}. The resulting constrain on the DP parameter, $R_{0}$ with a colored spectrum, based on the XENONnT data, becomes $R_{0} {\gtrsim ~ 10^{-14}\text{m}},$ i.e., several orders of magnitude weaker constraint compared to $R_{0}\sim 0.5 \text{nm}$ for white-noise models.

\section{Bounds on Collapse Models from Witnessing Entanglement at Short Times}
\label{app:C}

We consider two quantum oscillators initialized in a motional squeezed thermal state, with a squeezing parameter $r\in \mathbb{R}$ and initial motional temperature $\text{T}.$ The initial joint covariance matrix reads:

\begin{align}
    \sigma^{(0)}=\begin{pmatrix}
        \mathcal{N}e^{r} & 0 & 0 & 0\\
        0 & \mathcal{N}e^{-r} & 0 &0\\
        0 & 0& \mathcal{N}e^{r} & 0 \\
        0 & 0 & 0 & \mathcal{N}e^{-r}
    \end{pmatrix},
\end{align}
{where} $\mathcal{N} = \coth(\frac{\hbar\omega}{2k_{\text{b}}\text{T}}).$ For a bipartite Gaussian state with covariance matrix $\sigma$, a widely used entanglement measure is {logarithmic negativity}, defined as:
\begin{equation}
E_{\mathcal{N}} = \max\{0, -2\log_2 \tilde{\nu}_{-}\},
\end{equation}
where $\tilde{\nu}_{-}$ is the smallest symplectic eigenvalue of the 
partially transposed covariance matrix $\tilde{\sigma}$.

Given a $4\times 4$ covariance matrix in block form:
\begin{equation}
\sigma = \begin{pmatrix}
A & C \\
C^{\top} & B
\end{pmatrix},
\end{equation}
where $A$ and $B$ describe the local modes and $C$ encodes correlations, 
the symplectic eigenvalues of the partially transposed state are given by~\cite{Serafini2017}:
\begin{equation}
\tilde{\nu}_{\mp} = \sqrt{\frac{\tilde{\Delta} \mp \sqrt{\tilde{\Delta}^2 - 4 \det \sigma}}{2}},
\end{equation}
with
\begin{equation}
\tilde{\Delta} = \det A + \det B - 2 \det C.
\end{equation}

Finally, the logarithmic negativity is computed as:
\begin{equation}
E_{\mathcal{N}} = \max \{ 0, -2\log_2 \tilde{\nu}_{-} \}.
\end{equation}

\noindent
Evidently, a state is entangled if and only if $\tilde{\nu}_{-} < \frac{1}{2}$.

In our case, we compute the time-evolved covariance matrix, $\sigma(t),$ for Coulomb-coupled identical quantum oscillators with mass $m$ and frequency $\omega$, where the coupling is characterized by the frequency shift parameter, $\delta^{2}$. We also account for diffusive spontaneous collapse noise, with local noise coefficients $\mathcal{D}^{11}_{D}=\mathcal{D}^{22}_{D}$ and non-local noise coefficients $\mathcal{D}^{12}_{D}=\mathcal{D}^{21}_{D}$, with $D\in\{\text{CSL}, \text{DP}\}$ as well as the dissipative thermal noise, with dissipation coefficient $\gamma$ and thermal diffusion $\mathcal{D}_{\text{th}}.$ We find, upto $\mathcal{O}(t)$:

\begin{align}
    \tilde{\nu}_{-} &= \frac{1}{2}\left(\mathcal{N} + \frac{t e^{r}}{2m\hbar\omega}\left[\left(2\mathcal{D}^{11}_{D} + 2\mathcal{D}_{th} - 2 e^{-r}m\gamma\hbar\omega\mathcal{N}\right)-\sqrt{4(\mathcal{D}^{12}_{D})^{2}+\delta^{4}m^{2}\hbar^{2}\omega^{4}\mathcal{N}^{2}}\right]\right).
\end{align}

Here, $\mathcal{D}_{\text{th}} = 2mk_{\text{b}}\text{T}\gamma$, $\text{T}$ corresponds to environmental temperature, for $k_{\text{b}}\text{T} \gg \hbar\omega$, we can ignore the $\gamma$ dependent term in $\tilde{\nu}_{\pm}.$ 

The strongest bound on the collapse-induced diffusion is obtained when the two quantum oscillators are initialized at zero temperatures, i.e., $\mathcal{N} =1$; in this case, the condition to observe entanglement becomes ($\gamma \ll 1$):

\begin{align}\label{entanglement_cond}
    4\left(\mathcal{D}^{11}_{D} + \mathcal{D}_{\text{th}}\right)^{2} < 4(\mathcal{D}^{12}_{D})^{2} + \delta^{4}m^{2}\omega^{4}\hbar^{2}.
\end{align}

\subsection{Bounds on The CSL Model}

Using the diffusion coefficients derived for CSL, Eq.~\eqref{Dcsl_11} and Eq.~\eqref{CSL_correlating_diffusion}, we can bound $\lambda_{\text{CSL}}$ from the above entanglement observation criteria~\eqref{entanglement_cond}. To gain some intuition, we look at the case where $d\gg r_{\text{c}}$. In this regime, we can ignore $\mathcal{D}^{12}_{\text{CSL}}$ and  obtain a simplified expression for the bound:

\begin{align}\label{entanglement_bound}
    \mathcal{D}^{11}_{\text{CSL}} &<\frac{\delta^{2}m\omega^{2}\hbar\left(1-f\right)}{2}, \hspace{0.5cm} f= \frac{2\mathcal{ D}_{\text{th}}}{\delta^{2}m\omega^{2}\hbar}.
\end{align}

For consistency, we require $f <1$, which is precisely the condition on $\mathcal{D}_{\text{th}}$ to allow for Coulomb-mediated entanglement, in the absence of CSL. This requirement translates to:

\begin{align}
    Q &>\frac{4}{\delta^2}\underbrace{\frac{k_{\text{b}}\text{T}}{\hbar\omega}}_{n_{\text{th}}}.
\end{align}

It is interesting to note that the condition for interaction-induced entanglement is independent of the mass of the quantum particles, and only depends upon the ratio of thermal excitations, $n_{\text{th}}$, and the frequency shift, $\delta^2.$ In particular, higher thermal excitation relative to frequency shift is worse for entanglement generation.

It is clear from Eq.~\eqref{entanglement_bound} that the strongest bounds are obtained for weak coupling. With the parameters chosen to satisfy the condition $f < 1,$ specifically, using the parameters in the Table~\eqref{Table_2} and Table~\eqref{Table3}, we find, at $r_{\text{CSL}} = 10^{-7}\text{m}$, we can constrain $\lambda_{\text{CSL}} < 1.7\times 10^{-12}\text{s}^{-1}$ at $\text{T} = 1\text{K}.$ With more ambitious environmental temperatures, and detection of weaker entanglement, induced by smaller $\delta^2$, it is possible to probe $\lambda_{\text{CSL}} < 4\times10^{-14}\text{s}^{-1}$ (see Table~\eqref{Table3}). 

Although these limits are several orders of magnitude less than the most recent X-ray bound, we show that the X-ray bounds become significantly weaker for coloured extensions of CSL (see~\eqref{Xray-cCSL}); in contrast, the bounds obtained with an entanglement test will remain robust against such modifications. The bounds presented here makes use of the full entanglement witness criterion $\eqref{entanglement_cond}.$

\begin{table}[htbp]
\caption{Parameters used to constrain $\lambda_{\mathrm{CSL}}$ via witnessing entanglement.}\label{Table_2}
\centering
\begin{ruledtabular}
\begin{tabular}{lc}
\textbf{Parameter} & \textbf{Value} \\
\hline
Charge ($q$) & $150 \times 10^{-19}~\mathrm{C}$ \\
Mass ($m$) & $1.8 \times 10^{-17}~\mathrm{kg}$ \\
Radius ($R$) & $150~\mathrm{nm}$ \\
Trap frequency ($\omega$) & $2\pi \times 10^{3}~\mathrm{Hz}$ \\
Mechanical quality factor ($Q$) & $10^{11}$ \\
\end{tabular}
\end{ruledtabular}
\end{table}

\begin{table}[h!]
\centering
\caption{Bounds on $\lambda_{\text{CSL}}$ at $r_{\text{CSL}} = 10^{-7}\text{m}$ for different environmental temperatures, {$\text{T}$ }, mean separation, $d$ and corresponding frequency shift, $\delta^2$ and $f-$values. All other parameters are the same as in Table~\eqref{Table_2} 
}
\label{Table3}
\begin{ruledtabular}
\begin{tabular}{cccccc}
$\text{Choice}$ & $\lambda_{\text{CSL}}$(s$^{-1}$)&  $\text{T}$ & $d(\text{m})$ & $\delta^2$ & f \\
\hline
\text{A}& $1.7\times 10^{-12}$ & $1\text{K}$   & $2.0\times 10^{-4}$ & $1.4\times 10^{-3} $ & 0.58 \\
\text{B} & $1.19\times 10^{-13}$ & $100~\mathrm{mK}$   & $4.5.0\times 10^{-4}$ & $1.25\times 10^{-4} $ & 0.67 \\
\text{C} & $4.03\times 10^{-14}$ & $1~\mathrm{m K}$ & $8.0\times 10^{-4}$ & $2.2 \times10^{-5}$ & 0.37\\
\end{tabular}
\end{ruledtabular}
\label{tab:lambdaCSL}
\end{table}

\begin{figure}[H]
    \centering
    \includegraphics[width=0.6\linewidth]{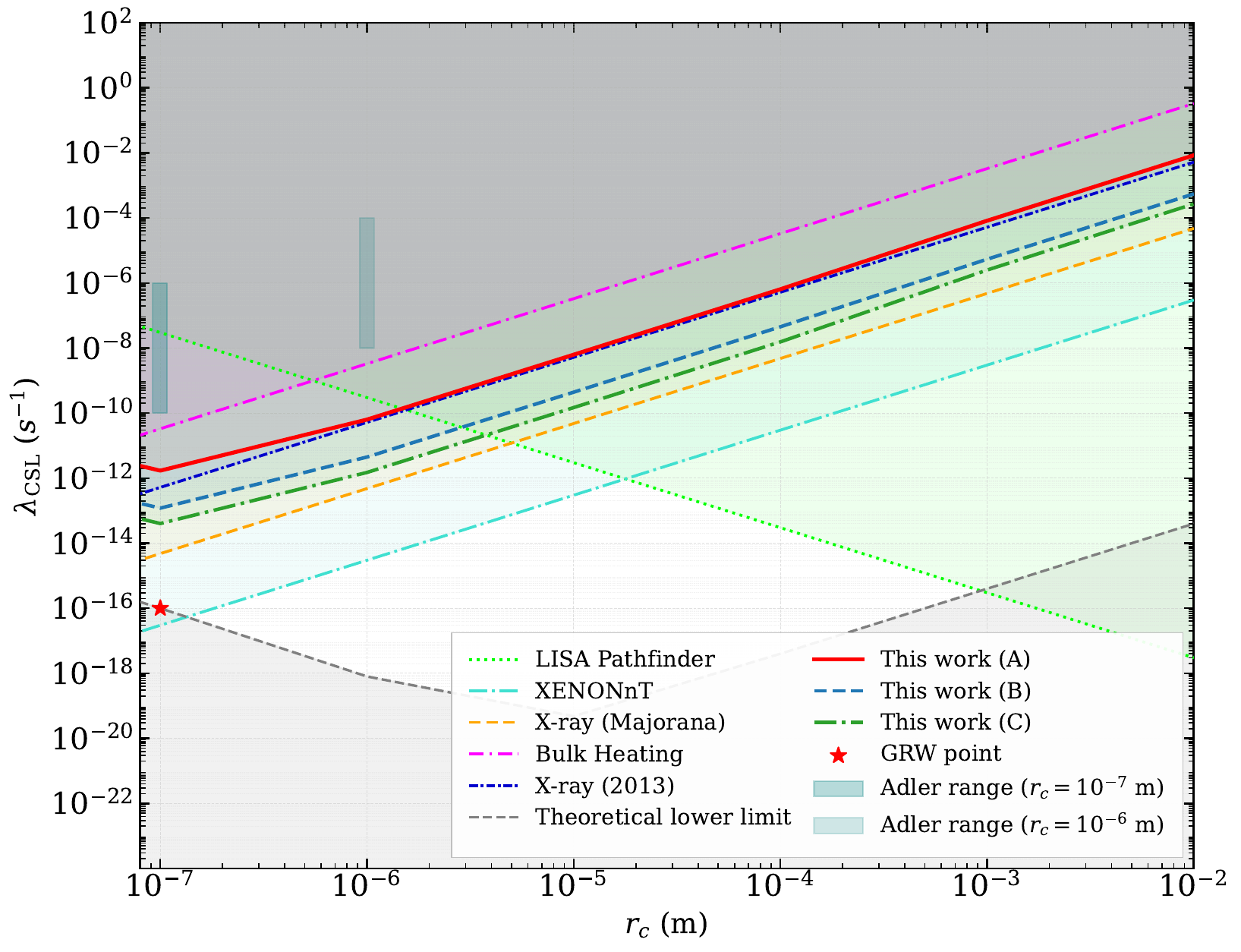}
    \caption{Our bounds at choices of experimental parameters in~\ref{Table3} considering a short-time Coulomb entanglement test of collapse models.}
    \label{fig:entanglement_bound}
\end{figure}

\subsection{Bounds on DP Model}\label{dp_entanglement_bound}

Unlike the CSL correlating noise, the DP correlating noise is long-ranged, in the regime, $d \gg \mathcal{R}_{D}$, $\mathcal{D}^{12}_{DP}$ $\rightarrow -\frac{G\hbar m^2}{d^3}$; which exactly the Newtonian-potential, upto a factor of $\hbar.$ Nevertheless, in this regime, we can ignore the correlating noise term compared to the local noise, which is unaffected by the large $d$ limit.

Proceeding as before, we have:

\begin{align}
    \mathcal{D}^{11}_{\text{DP}} < \frac{\delta^{2}m\omega^{2}\hbar\left(1-f\right)}{2}, \hspace{0.5cm} f= \frac{2\mathcal{ D}_{\text{th}}}{\delta^{2}m\omega^{2}\hbar}.
\end{align}

Using Eq.~\eqref{DP_coeff_appendix}, we find that in order to non-trivially constrain $R_{0}$, we require:

\begin{align}
    \gamma_{G} = \frac{Gm}{12\sqrt{\pi}\delta^{2}\omega^{2}} > R^{3}.
\end{align}

We find that, with $\rho = 2.6\times10^{3}\text{Kg}{m}^{-3}$, $\omega = 2\pi\times 100\text{Hz}$, we need to observe entanglement caused by  $\delta^{2} < 8.6\times 10^{-14}$ which is extremely tiny! i.e., the Diosi-Penrose model, in most cases, allows for the generation of static-Coulomb mediated entanglement between macroscopic charged systems.

\section{Dynamics with Diosi-Penrose Collapse}\label{app:D}

In this section, we explore the Coulomb-mediated reduction in variance below the thermal level with the DP model; the DP diffusion coefficients for a macroscopic rigid sphere are derived in Eq.~\eqref{DP_coeff_appendix}.  The diffusion coefficients are reproduced below for convenience:

\begin{align}\label{DP_coeff}
    \mathcal{D}^{11}_{\text{DP}}&=\mathcal{D}^{22}_{\text{DP}} \approx \frac{G\hbar m^{2}}{12\sqrt{\pi}\mathcal{R}^{3}_{D}},\\
    \mathcal{D}^{12}_{\text{DP}} &= \mathcal{D}^{21}_{\text{DP}} \approx\frac{G\hbar m^{2}}{4}\left(\frac{(4 \mathcal{R}^{2}_{D} + d^{2})e^{-\frac{d^{2}}{4 \mathcal{R}^{2}_{D}}}}{\sqrt{\pi}\mathcal{R}^{3}_{D}d^{2}} - \frac{4\text{Erf}(\frac{d}{2{\mathcal{R}_{D}}})}{d^{3}}\right).\label{DP_12}
\end{align} 

Here, we have defined $\mathcal{R}_{D} = \sqrt{R^{2}+R^{2}_{0}}$, where $R$ is the size of the rigid sphere, $R_{0}$ is the DP parameter, and $d$ is the mean distance between the COM of the two masses. Note that here the effective regulator is $\mathcal{R}_{D}$ and the linearized collapse dynamics remain valid as long as the center-of-mass motion occurs on a length scale much smaller than $\mathcal{R}_{D}.$ 

Following a similar computation with $\mathcal{D}^{d}_{\text{DP}}$ \textit{replacing} $\mathcal{D}^{d}_{\text{CSL}}$, we find that observing a squeezed thermal state imposes a non-trivial constraint on $R_{0}$ only when:

\begin{align}\label{eta_{G}}
     \eta_{G} = \frac{G \rho\sqrt{\pi}}{{9\delta^{2}}\omega\gamma \mathcal{N} } > 1.
\end{align}

With the constrain on $R_{0}$ satisfying: $R^{2}_{0} > \left(\eta_{G}^{2/3}-1\right)R^{2}.$ Achieving $\eta_{G} > 1$ requires meeting the outstanding challenge of resolving squeezing caused by an extremely low-frequency shift and achieving a steady state with extremely low motional temperature. Considering a mean separation, $d = 6.5\times 10^{-3}\text{m}$, a trap frequency $\omega = 2\pi\times1.28\text{KHz}$ and $q = 300\text{e}$ yields $\delta^{2} = 1.26 \times 10^{-7}$. With a steady state temperature of $100\text{nK},$ mechanical damping $\gamma = 8\times 10^{-6}\text{s}^{-1},$ and a density, $\rho= 2.6\times10^{3}\text{Kg}\text{m}^{-3}$ we can satisfy $\eta_{G} > 1$ and constrain $R_{0} = 5.9 \times 10^{-8}\text{m}$, which is an improvement of almost two orders of magnitude to the current best X-ray based bound on $R_{0} \sim 0.5 \text{nm}.$ Achieving the required experimental conditions is admittedly unfeasible. The key challenge here is that the effective DP parameter becomes $\mathcal{R}_{D}$ rather than $R_{0}$; as a result, isolating the contribution of $R_{0}$ becomes challenging. Evidently, this issue is generic to testing the Diosi-Penrose model with macroscopic systems and unrelated to the specific experimental scheme.

\end{document}